\documentclass[reprint,superscriptaddress,showkeys,amsmath,amssymb,aps]{revtex4-2}

\usepackage{graphicx}
\usepackage{dcolumn}
\usepackage{bm}
\usepackage[hidelinks]{hyperref}
\usepackage{siunitx}
\usepackage{comment}
\usepackage{esvect}
\usepackage{gensymb}
\usepackage{textcomp}
\usepackage{xcolor}

\begin{document}

%\preprint{APS/123-QED}

%\title{Low dissipation macroscopic levitating rotors at room temperature}% Force line breaks with \\
%\title{Hybrid diamagnetically levitated nanomechanical resonators for magnetic field sensing}% Force line breaks with 
\title{Levitated nano-trampoline resonators for magnetic field sensing}
%\\
%\begin{comment}
\author{Xianfeng Chen}
\email{xianfeng\_chen@a-star.edu.sg}
\affiliation{A*STAR Quantum Innovation Centre(Q.InC), Agency for Science, Technology and Research(A*STAR), 2 Fusionopolis Way, 08-03 Innovis 138634, Singapore}
%\affiliation{Institute for Materials Research and Engineering(IMRE), Agency for Science, Technology and Research(A*STAR), 2 Fusionopolis Way, 08-03 Innovis 138634, Singapore}
%\affiliation{Centre for Quantum Technologies, National University of
%Singapore, Singapore, 117543, Singapore}

\author{Nirmala Raj}%
\affiliation{A*STAR Quantum Innovation Centre(Q.InC), Agency for Science, Technology and Research(A*STAR), 2 Fusionopolis Way, 08-03 Innovis 138634, Singapore}
%\affiliation{Institute for Materials Research and Engineering(IMRE), Agency for Science, Technology and Research(A*STAR), 2 Fusionopolis Way, 08-03 Innovis 138634, Singapore}

\author{Matthew R. Chua}%
\affiliation{Institute for Materials Research and Engineering(IMRE), Agency for Science, Technology and Research(A*STAR), 2 Fusionopolis Way, 08-03 Innovis 138634, Singapore}

\author{Yi Fan Chen}%
\affiliation{Institute for Materials Research and Engineering(IMRE), Agency for Science, Technology and Research(A*STAR), 2 Fusionopolis Way, 08-03 Innovis 138634, Singapore}

\author{Chenyue Gu}%
\affiliation{A*STAR Quantum Innovation Centre(Q.InC), Agency for Science, Technology and Research(A*STAR), 2 Fusionopolis Way, 08-03 Innovis 138634, Singapore}
\affiliation{Centre for Quantum Technologies, National University of
Singapore, Singapore, 117543, Singapore}
\affiliation{Department of Physics, National University of Singapore, 2 Science Drive 3, 117542, Singapore}

\author{Minxing Xu}
\affiliation{A*STAR Quantum Innovation Centre(Q.InC), Agency for Science, Technology and Research(A*STAR), 2 Fusionopolis Way, 08-03 Innovis 138634, Singapore}
%\affiliation{Institute for Materials Research and Engineering(IMRE), Agency for Science, Technology and Research(A*STAR), 2 Fusionopolis Way, 08-03 Innovis 138634, Singapore}

\author{Young-Wook Cho}
\affiliation{A*STAR Quantum Innovation Centre(Q.InC), Agency for Science, Technology and Research(A*STAR), 2 Fusionopolis Way, 08-03 Innovis 138634, Singapore}
%\affiliation{Institute for Materials Research and Engineering(IMRE), Agency for Science, Technology and Research(A*STAR), 2 Fusionopolis Way, 08-03 Innovis 138634, Singapore}
\affiliation{Centre for Quantum Technologies, National University of
Singapore, Singapore, 117543, Singapore}

\author{Syed M. Assad}
\affiliation{A*STAR Quantum Innovation Centre(Q.InC), Agency for Science, Technology and Research(A*STAR), 2 Fusionopolis Way, 08-03 Innovis 138634, Singapore}
%\affiliation{Institute for Materials Research and Engineering(IMRE), Agency for Science, Technology and Research(A*STAR), 2 Fusionopolis Way, 08-03 Innovis 138634, Singapore}

\author{Lu Ding}%
\email{dinglu81@gmail.com}
\affiliation{Institute for Materials Research and Engineering(IMRE), Agency for Science, Technology and Research(A*STAR), 2 Fusionopolis Way, 08-03 Innovis 138634, Singapore}

\author{Ping Koy Lam}
\email{pingkoy@a-star.edu.sg}
\affiliation{A*STAR Quantum Innovation Centre(Q.InC), Agency for Science, Technology and Research(A*STAR), 2 Fusionopolis Way, 08-03 Innovis 138634, Singapore}
%\affiliation{Institute for Materials Research and Engineering(IMRE), Agency for Science, Technology and Research(A*STAR), 2 Fusionopolis Way, 08-03 Innovis 138634, Singapore}
\affiliation{Centre for Quantum Technologies, National University of
Singapore, Singapore, 117543, Singapore}
%\affiliation{Centre for Quantum Technologies, National University of Singapore, Singapore 117543, Singapore}
%\end{comment}

\date{\today}% It is always \today, today,

%TC:ignore
\begin{abstract}
Levitated systems and high-$Q$ membrane nanomechanical resonators have achieved exceptional sensitivity in precision sensing, but functionalizing such resonators for practical applications without degrading their low dissipation remains challenging. Here, we combine diamagnetic levitation with a high-$Q$ nanomechanical resonator to realize a high-precision magnetometer {for sensing weak oscillating magnetic fields}. A macroscopic diamagnetically levitated graphite plate acts as a free-floating proof mass that couples strongly to magnetic fields, converting them into mechanical motion that is resonantly amplified by a low-dissipation nano-trampoline resonator. Operating at room temperature and without magnetic shielding, we achieve a peak magnetic field sensitivity of \SI{4.5}{pT/\sqrt{Hz}} for a resonator with a mechanical $Q$ of \si{6\times10^6} at \SI{443}{kHz}. The system sensitivity is limited by thermomechanical noise. With further improvements in mechanical $Q$, this hybrid levitated platform offers a pathway toward femtotesla-level {AC} magnetic field sensing, establishing diamagnetically levitated nanomechanical resonators as a new class of high-sensitivity magnetometers at room temperature. 
\end{abstract}
%TC:endignore

%Use showkeys class option if keyword
\maketitle

%\tableofcontents

\section{Introduction}
The detection of extremely weak forces is a central challenge in precision measurement, with broad relevance to applications ranging from biomedical imaging and navigation to spin detection and tests of fundamental physics~\cite{abramovici1992ligo,rugar2004single,arash2015review,mason2019continuous,barzanjeh2022optomechanics}. Nanomechanical resonators are among the most sensitive force transducers due to their small effective mass and high resonance frequencies \cite{moser2013ultrasensitive,gavartin2012hybrid}. Their ultimate sensitivity is fundamentally constrained by coupling to the thermal bath through dissipation \cite{ekinci2004ultimate}, which in high vacuum is typically dominated by clamping and support-induced losses.

Recent advances in dissipation dilution have enabled high-stress nanomechanical resonators to reach exceptional performance, with $Q$-factors exceeding \si{10^9}~\cite{beccari2022strained,cupertino2024centimeter,engelsen2024ultrahigh}. For sensing applications, however, nanomechanical resonators must couple to external signals, which generally requires functionalization. In magnetic field sensing, which is central to many applications ranging from biomedical imaging to mineral exploration \cite{edelstein2007advances}, this coupling is commonly achieved by employing magnetostrictive materials \cite{forstner2012cavity,PhysRevLett.125.147201,xu2024subpicotesla} or by attaching additional magnetic elements \cite{fischer2019spin,zhang2025membrane,gottardo2025silicon}. Such approaches inevitably constrain the material options, degrade $Q$-factors, and increase fabrication complexity, limiting the achievable sensitivity.

Levitation offers an alternative route to eliminate clamping loss and thereby suppress dissipation~\cite{gonzalez2021levitodynamics}. Optical, electric, and superconducting levitation techniques have demonstrated exquisite isolation and ultra-low dissipation \cite{delic2020cooling,hofer2023high,dania2024ultrahigh,fuchs2024measuring,ahn2020ultrasensitive}, but typically rely on complex trapping schemes, small levitated masses, or cryogenic environments. Diamagnetic levitation, by contrast, provides passive and stable levitation of macroscopic objects at room temperature and without active feedback \cite{simon2000diamagnetic,chen2020rigid,yin2022experiments,tian2024feedback,chen2024nonlinear}. Owing to its strong coupling to external fields and forces, diamagnetic levitation has enabled precision sensors including force sensors \cite{yin2022experiments,tian2025constraints}, accelerometers \cite{garmire2007diamagnetically,chen2022diamagnetic,wang2025ultra}, gravimeters \cite{leng2024measurement,wang2020diamagnetic}, and gyroscopes \cite{geim2001detection,chen2026levitated}. However, the sensing performance of purely diamagnetically levitated systems has thus far been restricted to low frequencies due to their intrinsically low mechanical resonance frequencies.

Here we introduce a hybrid levitated nanomechanical resonator that combines a macroscopic diamagnetically levitated graphite plate with a high-stress silicon nitride membrane, forming a free-floating, low-dissipation mechanical system at room temperature. In this architecture, the levitated graphite plate acts as a macroscopic proof mass that couples directly to magnetic fields and converts them into mechanical forces, while the silicon nitride membrane serves as a high-$Q$ nanomechanical amplifier and readout element. Importantly, the motion of the levitated graphite alone is too weak to be directly detected at frequencies far above its mechanical resonances, whereas the nanomechanical resonator by itself exhibits no direct response to magnetic fields. Their hybridization overcomes both limitations, enabling efficient transduction of {AC} magnetic fields into amplified nanomechanical motion without employing magnetic materials or introducing additional functional elements on the resonator.

%Here we introduce a hybrid levitated nanomechanical resonator that combines a macroscopic diamagnetically levitated graphite plate with a high-stress silicon nitride (Si$_3$N$_4$) membrane, forming a free-floating, low-dissipation mechanical system at room temperature. The levitated graphite plate acts as a macroscopic proof mass that converts magnetic fields into mechanical forces, while the silicon nitride membrane serves as a high-$Q$ nanomechanical amplifier and readout element. This hybridization enables efficient transduction of magnetic fields into amplified nanomechanical motion without employing magnetic materials or introducing additional functional elements on the nanomechanical resonator. Operating at room temperature and under Earth's magnetic field, we realize a magnetic field sensor based on this platform and achieve a calibrated sensitivity of \SI{4.5}{pT/\sqrt{Hz}}, approaching the thermomechanical noise limit. 
%These results establish hybrid diamagnetically levitated nanomechanical resonators as a new class of force sensors that combine macroscopic levitation with nanoscale mechanical coherence, opening a pathway toward ultralow-dissipation precision sensing under ambient conditions.

\section{Results}
\subsection{Working principle of the hybrid platform}
\begin{figure}[!t]
\centering
\includegraphics[width=8cm]{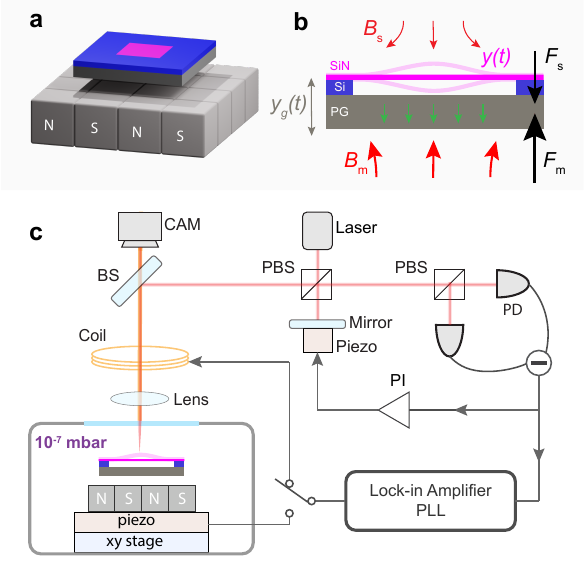}
\caption{\label{fig:concept} Hybrid diamagnetically levitated nanomechanical magnetometer. (a) Schematic of a pyrolytic graphite (PG) plate levitating above an array of permanent magnets with alternating magnetization, where N and S represent the north and south poles of the magnet, respectively. A chip consisting of a Si substrate and a nanometer-thick SiN square membrane is placed on top of the graphite. (b) Cross-section of the chip (not to scale). The green arrows represent the diamagnetic response of the graphite. 
%The red arrows illustrate the magnetic field generated by the NdFeB magnets $B_\mathrm{m}$ and an externally applied rf magnetic field {amplitude} $B_\mathrm{s}$; the green arrows represent the diamagnetic response of the graphite; the black arrows stand for the resulting magnetic forces acting on the levitated proof mass from the magnets ($F_\mathrm{m}$) and magnetic signal ($F_\mathrm{s}$), respectively; $y_\mathrm{g}(t)$ and $y(t)$ represent the motion of the graphite plate and the membrane, respectively.
(c) Experimental setup for magnetic field sensing, including optical interferometric readout of the nanomechanical resonator motion (see also SI-IV).
}
\end{figure}
The hybrid levitated nanomechanical magnetometer is illustrated in Fig.~\ref{fig:concept}. It consists of a \(10.2\times10.2\times\SI{1.2}{mm^{3}}\) pyrolytic graphite (PG) plate levitated above an array of \SI{4}{mm} NdFeB cube magnets with alternating magnetization, together with square nanomechanical membrane resonators fabricated from \SI{100}{nm}-thick high-stress silicon nitride films (SiN), with side lengths ranging from \SI{0.2}{mm} to \SI{1.4}{mm} (see Fig.~\ref{fig:concept}a,b and SI-I). Owing to the strong diamagnetism of pyrolytic graphite~\cite{chen2021diamagnetically,chen2022diamagnetic}, the graphite and the stacked Si/SiN chip can be stably levitated in a three-dimensional magnet-gravitational trap without any power supply.

As illustrated in Fig.~\ref{fig:concept}b, the diamagnetically levitated graphite plate experiences three main forces: the magnetic force arising from the permanent magnets, \(\mathbf{F}_{\mathrm{m}}\); the magnetic force {amplitude} associated with the external signal field {amplitude}, \(\mathbf{F}_{\mathrm{s}}\); and the gravitational force acting on both the graphite plate and the SiN chip (not shown). The total magnetic force per unit volume acting on the graphite can be expressed as $\mathbf{F} = \mathbf{F}_{\mathrm{m}}+\mathbf{F}_{\mathrm{s}}  =\chi/\mu_{0}\,\mathbf{B}\nabla \mathbf{B},$ where \(\chi\) denotes the magnetic susceptibility of graphite; \(\mu_{0}\) is the vacuum permeability; and \(\mathbf{B} = \mathbf{B}_{\mathrm{m}} + \mathbf{B}_{\mathrm{s}}\) is the total magnetic field with contributions from the permanent magnets, \(\mathbf{B}_{\mathrm{m}}\), and the external signal field {amplitude}, \(\mathbf{B}_{\mathrm{s}}\). Since \(|B_{\mathrm{m}}| \gg |B_{\mathrm{s}}|\), the static levitation force is dominated by the magnetic field of the permanent magnets, \(\mathbf{F}_{\mathrm{m}} = (\chi/\mu_{0}) \mathbf{B}_{\mathrm{m}}\nabla \mathbf{B}_{\mathrm{m}}\), which balances gravity at the equilibrium levitation height.

When an external {AC} magnetic signal is applied, a weak perturbative force amplitude \(\mathbf{F}_{\mathrm{s}}\) is generated, driving the graphite plate into motion $y_\mathrm{g}$ (Fig.~\ref{fig:concept}b). Although both the magnetic field amplitude \(\mathbf{B}_{\mathrm{s}}\) and the field gradient \(\nabla \mathbf{B}_{\mathrm{s}}\) contribute to the magnetic force, the dominant contribution to the perturbative force in the present experiment arises from the product of the signal magnetic field {amplitude} \(\mathbf{B}_{\mathrm{s}}\) and the static magnetic field gradient of the permanent magnets \(\nabla \mathbf{B}_{\mathrm{m}}\), such that (see also SI-II and SI-VII):
\begin{equation}
  \mathbf{F}_{\mathrm{s}} \approx \frac{\chi}{\mu_{0}} \mathbf{B}_{\mathrm{s}} \nabla \mathbf{B}_{\mathrm{m}}.  
  \label{eq:Fs}
\end{equation}
Consequently, the device operates as a magnetic field sensor, and is calibrated accordingly.

\begin{figure*}[!t]
\centering
\includegraphics[width=17cm]{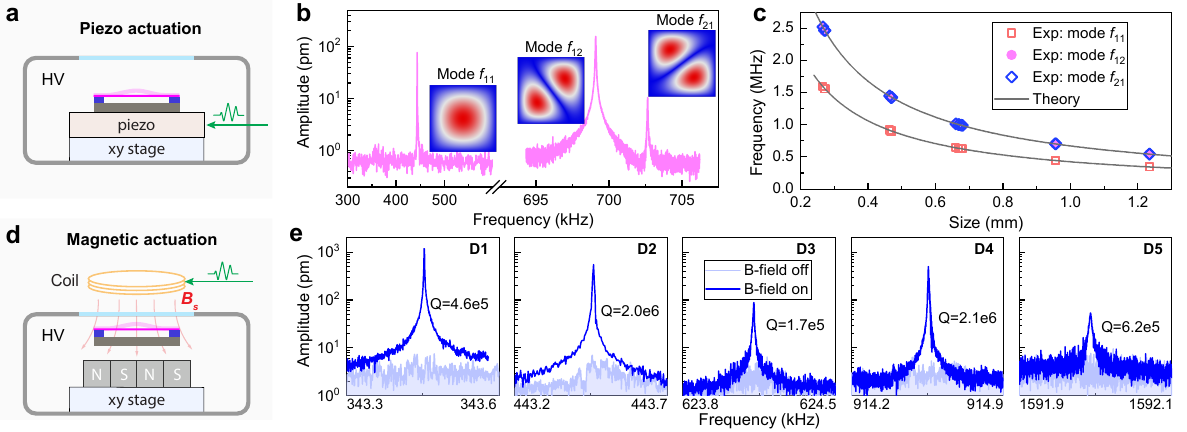}
\caption{\label{fig:B-filed response} Dynamic characterization and magnetic response of the hybrid levitated nanomechanical resonator. (a) Schematic of mechanical actuation using a piezoelectric disk. HV: high-vacuum chamber. (b) Measured frequency response of a representative nanomechanical resonator under piezoelectric actuation, showing the first three mechanical modes; the corresponding simulated mode shapes are shown next to each peak. (c) Resonance frequencies of the first three modes for square membrane resonators with different side lengths, measured via piezoelectric actuation. (d) Schematic of magnetic actuation using a coil. (e) Frequency response near the first resonance of five representative devices with magnetic actuation switched on and off. Quality factors extracted from ringdown measurements are indicated next to the corresponding resonance peaks (see SI-VI).}
\end{figure*}

The motion of the levitated graphite plate is mechanically transferred to the nanomechanical membrane resonator, which acts as a high-\(Q\) mechanical amplifier (Fig.~\ref{fig:concept}b). The motion of the nanomechanical resonator can be modeled as:
\begin{equation}
    %\left\{
    \begin{aligned}      
    %\Ddot{y}+\frac{\omega_\mathrm{pg}}{Q_\mathrm{pg}}\Dot{y}+\omega_\mathrm{pg}^2 y&=\frac{F_\mathrm{sg}}{m_\mathrm{\mathrm{pg}}}\cos(\omega t), \\
    \Ddot{y}+\frac{\omega_0}{Q}\Dot{y}+\omega_0^2(y-y_\mathrm{g})&=\frac{F_\mathrm{th}}{m_\mathrm{eff}},
    \end{aligned}
    %\right.
\end{equation}
%where $m_\mathrm{pg}, Q_\mathrm{pg}$, and $\omega_\mathrm{pg}$ are the effective mass, $Q$-factor, and resonance frequency of the levitated graphite resonator; 
where $y$, $m_\mathrm{eff}$, $Q$, and $\omega_\mathrm{0}$ are the displacement, effective mass, $Q$-factor, and resonance frequency of the nanomechanical resonator; $F_\mathrm{th}$ is the force due to thermomechanical noise. The nanomechanical resonator's motion is driven by the graphite's motion $y_\mathrm{g}$. {At radio frequencies, the direct motion of the levitated graphite plate is strongly suppressed, while the high-$Q$ nanomechanical resonator provides resonant enhancement of weak magnetic signals.}

%At radio frequencies ($>\SI{100}{kHz}$), the direct mechanical response of the levitated graphite plate $y_\mathrm{g}$ is strongly suppressed, scaling as $1/\omega^2$ ($\omega$ is the driving frequency), causing the resulting motion to fall below the detection imprecision. By contrast, owing to the high mechanical quality factors of the nanomechanical resonators, even extremely weak magnetic fields give rise to a pronounced resonant response. 

The resulting motion of the membrane resonator is detected using an optical interferometer with balanced detection, as shown in Fig.~\ref{fig:concept}c (see SI-IV for experimental details). All measurements are performed in a vacuum chamber at pressures below \SI{5e-7}{mbar} to suppress air damping. This separation of force transduction and mechanical resonance enhancement enables strong magnetic coupling without introducing additional effects into the nanomechanical resonator, and the magnetic field sensitivity of the levitated magnetometer is calibrated using an electromagnetic coil that generates a well-defined magnetic field.

\subsection{Dynamics characterization of nano-trampoline resonators}

To evaluate the magnetic sensing performance of the levitated nanomechanical resonators, we first characterize their resonance frequencies and mechanical quality factors using piezoelectric actuation (Fig.~\ref{fig:B-filed response}a). We place the SiN chip directly on a piezoelectric disk, then drive it over a broad frequency range to excite the trampoline resonators and identify their mechanical modes. Figure~\ref{fig:B-filed response}b shows a representative frequency response, revealing the first three resonant modes of a square silicon nitride membrane with a side length of \SI{~460}{\um}. The corresponding mode shapes simulated using COMSOL Multiphysics are shown alongside the measured peaks.

For a high-stress square SiN membrane, the resonance frequencies are given by~\cite{schmid2016fundamentals}:
\begin{equation}
    f_{mn} = \frac{1}{2L}\sqrt{\frac{\sigma_{\mathrm{t}}}{\rho}}\sqrt{m^{2}+n^{2}}, \quad m,n=1, 2, 3, ...
    \label{eq:frequency}
\end{equation}
where \(L\), \(\sigma_{\mathrm{t}}\), and \(\rho\) denote the membrane side length, tensile stress, and material density, respectively. From Fig.~\ref{fig:B-filed response}b, the ratios of $f_{12}/f_{11}$ and $f_{21}/f_{11}$ are approximately 1.58, in good agreement with the analytical expectation. 
%The small splitting between $f_{21}$ and $f_{12}$ is attributed to the fabrication-induced imperfection from a perfect square geometry. 
To further verify the membrane quality and the scaling of the resonance frequencies, we fabricate resonators with different side lengths and measure their first three resonance modes. The extracted resonance frequencies are summarized in Fig.~\ref{fig:B-filed response}c. 
%Fitting the data using Eq.~(\ref{eq:frequency}) yields a tensile stress of \(\sigma_{\mathrm{t}}=\SI{1.115}{GPa}\), consistent with the value specified by the supplier.

\begin{figure*}[!t]
\centering
\includegraphics[width=17cm]{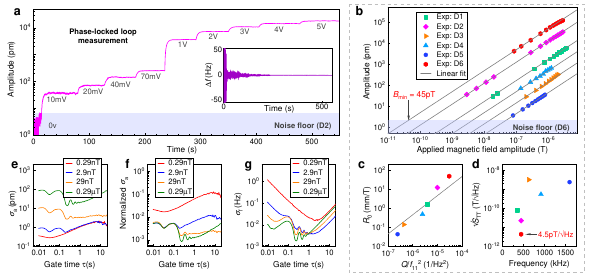}
\caption{\label{fig:step response}  Magnetic field sensing and sensitivity characterization. (a) Amplitude response of a representative nanomechanical resonator (device D2) to applied magnetic field {amplitudes}, obtained by using a phase-locked loop. The inset shows the corresponding resonance frequency fluctuations recorded over the same measurement duration. {The shaded gray region represents the measured noise floor of device D2.} (b) Measured peak amplitude as a function of applied magnetic field {amplitude} for six different devices. Error bars are smaller than the marker size. The shaded gray region indicates the measured noise floor of device D6. (c) Responsivity $\mathcal{R}_0$ {as a function of $Q/f_{11}^2$} for different devices, extracted from the linear fits in panel (b). (d) Magnetic field sensitivity $\sqrt{S_\mathrm{TT}}$ of the devices in panel (b) based on the measured noise floor. (e–g) Allan deviations calculated from the time traces of the resonant peak amplitude and resonance frequency over a \SI{5}{min} interval for device D6, driven at four different {magnetic field amplitudes}. In panel (f), the Allan deviation of the amplitude is normalized by the mean amplitude at the corresponding {magnetic field amplitude}.}
\end{figure*}

\subsection{Magnetic field response}
After characterizing the mechanical properties of the membrane resonators, we next investigate their response to magnetic excitation. We focus on the fundamental mode $f_{11}$ of each device and apply a time-varying magnetic field $B_\mathrm{s}\cos({wt})$ using an electromagnetic coil positioned on top of the vacuum chamber (Fig.~\ref{fig:B-filed response}d). The drive frequency $w$ is swept across the resonance frequencies while the resonator displacement amplitude is recorded. Figure~\ref{fig:B-filed response}e shows the frequency responses of five representative devices D1-5 with the magnetic coil switched on and off. In all cases, a pronounced resonance peak appears only when the magnetic drive is applied, confirming that the levitated nanomechanical resonators respond to external radio-frequency magnetic fields. {Since the magnets are rigidly attached to a massive support structure, the measured magnetic response is attributed predominantly to the levitated graphite plate (see SI-IX for more details).} By varying the size of the membranes, the magnetic response is observed over a wide frequency range, from \SI{300}{kHz} to \SI{1.6}{MHz}, highlighting the broadband operation enabled by the hybrid levitated architecture.

%\section{Response to magnetic force}

To quantitatively characterize the magnetic field response, we employ a phase-locked loop (PLL, see SI-V for details) to continuously track the resonance frequency $f_{11}$ of the nanomechanical resonator and record its displacement amplitude, while varying the voltage {amplitude} applied to the excitation coil. Figure~\ref{fig:step response}a shows the real-time amplitude response and frequency fluctuation (see the inset) of a representative device D2. As the applied voltage {amplitude} increases, the resonator amplitude increases accordingly. Owing to the high mechanical quality factor $Q=2\times10^6$, a ring-up transient is observed immediately following each voltage step.

We perform this step-response measurement for six different devices by varying the applied voltage {amplitude} from 0 to \SI{10}{V}. As shown in Eq.~(\ref{eq:Fs}), the magnetic actuation force {amplitude} is proportional to the applied magnetic field {amplitude} generated by the coil. Using a fluxgate magnetometer in combination with numerical simulations, we calibrate the coil and establish the relationship between the magnetic field {amplitude} and the applied voltage {amplitude} at different frequencies (see SI-VII). This calibration allows us to plot the steady-state amplitude response as a function of the applied magnetic field {amplitude}, as shown in Fig.~\ref{fig:step response}b. Linear fits to the low-amplitude data (first four lowest points for each device) are also superimposed. At higher drive amplitudes, deviations from linearity are observed, consistent with the onset of nonlinear dynamics commonly encountered in high-\(Q\) nanomechanical resonators~\cite{catalini2021modeling,li2024strain}. Nevertheless, the amplitude response remains linearly proportional to the applied magnetic field {amplitude} over a wide dynamic range, spanning approximately five orders of magnitude for device D6. The slope of each curve in Fig.~\ref{fig:step response}b defines the responsivity of the device at its resonance, $\mathcal{R}_0 = Y_0/{B_{\mathrm{s}}}$, which quantifies the displacement amplitude response $Y_0$ per unit magnetic field {amplitude}. The responsivities extracted from linear fits are summarized in Fig.~\ref{fig:step response}c. For levitated magnetometers based on high-stress square membranes, the responsivity follows the scaling relation \(\mathcal{R}_0 \propto Q/{f_{11}}^2\) (see SI-III), which generally aligns with our observation in Fig. \ref{fig:step response}c, {with deviation that is probably due to the model simplification and imperfect fabrication.}

\subsection{Sensitivity and stability}

\begin{figure*}[!t]
\centering
\includegraphics[width=16cm]{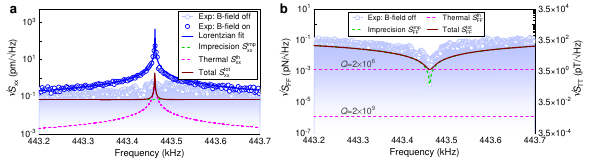}
\caption{\label{fig:sensitivity} Noise and sensitivity analysis. (a) Measured and calculated amplitude noise spectrum of device D2 around its resonance frequency. Total $S_\mathrm{xx}^{\mathrm{tot}}$ represents the PSD due to both $S_\mathrm{xx}^{\mathrm{th}}$ and $S_\mathrm{xx}^{\mathrm{imp}}$ (b) Equivalent force and magnetic field noise spectrum inferred from the data in panel (a) using the mechanical transfer function of the resonator and the measured Q-factor.}
\end{figure*}

Another key performance metric of a resonant sensor is its sensitivity, i.e., the minimum detectable magnetic field {amplitude} in our case. The sensitivity is determined by the measured noise floor, indicated in Fig.~\ref{fig:step response}b. For device D6, the minimum detectable magnetic field {amplitude} is \SI{45}{pT}. Knowing the measurement bandwidth (\(BW=\SI{100}{Hz}\) in our measurements), the magnetic field sensitivity in units of \si{T/\sqrt{Hz}} can be expressed as~\cite{PhysRevLett.125.147201}:
\begin{equation}
    \sqrt{S_{\mathrm{TT}}} = \frac{ B_{\mathrm{s}}}{\mathrm{SNR}_{\mathrm{a}}\sqrt{BW}}
    = \frac{ B_{\mathrm{min}}}{\sqrt{BW}},
    \label{eq:sensitivity}
\end{equation}
where \(\mathrm{SNR}_{\mathrm{a}}\) is the amplitude signal-to-noise ratio and \( B_{\mathrm{min}}\) denotes the minimum detectable magnetic field {amplitude}. From Eq.~(\ref{eq:sensitivity}) and the data in Fig.~\ref{fig:step response}b, we extract the sensitivities of all six devices and plot them versus resonance frequency in Fig.~\ref{fig:step response}d, achieving a best sensitivity of \SI{4.5}{pT/\sqrt{Hz}} {at the device's first resonance frequency}.

To quantify the stability of the levitated magnetometer, we record the displacement-amplitude response of device D6 over a \SI{5}{min} interval with the PLL while driving it at four different {magnetic field amplitudes}, and compute the corresponding Allan deviations (see SI-VIII), as shown in Fig.~\ref{fig:step response}e.
%Because a \SI{100}{Hz} low-pass filter is applied in the lock-in amplifier, only data with gate times \(\tau>\SI{0.01}{s}\) are considered. 
At low drive amplitude~({$\SI{0.29}{nT}$}), the Allan deviation exhibits behavior characteristic of a linear mechanical resonator, with \(\sigma_{\mathrm{a}}\propto\tau^{-1/2}\) at short integration times due to white noise. 
%At higher drive amplitudes, more complex behavior is observed, likely arising from Duffing-type nonlinearities~\cite{manzaneque2023resolution}. 
The normalized Allan deviation (Fig.~\ref{fig:step response}f) shows that relative amplitude fluctuations decrease with increasing drive amplitude, achieving a minimum fluctuation of \SI{0.1}{\percent} at $\tau\approx\SI{0.2}{s}$ when driven with {$\SI{0.29}{\micro T}$}. The resonance frequency stability is shown in the inset of Fig.~\ref{fig:step response}a and in Fig.~\ref{fig:step response}g, showing a better stability with higher drive.

{To identify the dominant sensitivity limits, we analyze the displacement noise spectrum of the levitated nanomechanical resonator, as shown in Fig.~\ref{fig:sensitivity}a. The measurement imprecision is obtained by positioning the interferometric laser spot on a non-undercut region of the SiN film.}
%To identify the factors limiting the sensitivity of the levitated nanomechanical magnetometer, we perform a detailed analysis of the dominant noise contributions, with particular emphasis on thermomechanical noise and measurement imprecision. The frequency-dependent displacement amplitude spectral densities, \(\sqrt{S_{xx}}\), associated with different noise sources are shown in Fig.~\ref{fig:sensitivity}a. The measurement imprecision (dashed green curve) is obtained by positioning the interferometric laser spot on a region of the SiN film without undercut, thereby capturing all detection-related noise contributions, including photon shot noise and electronic noise.
Near resonance, the measured displacement noise of the nanomechanical resonator exceeds the measurement imprecision, indicating that the resonator motion is dominated by thermomechanical fluctuations (see also Fig.~\ref{fig:B-filed response}e). The thermomechanical noise spectrum can be expressed as
$S_{xx}^{\mathrm{th}}(\omega)=|\chi(\omega)|^{2}S_{FF}^{\mathrm{th}}$,
where \(S_{FF}^{\mathrm{th}}=4k_{\mathrm{B}}T m_{\mathrm{eff}}/Q\) is the thermal force noise spectral density, and \(\chi(\omega) = [m_{\mathrm{eff}}(\omega_{0}^{2}-\omega^{2}+i\omega_{0}\omega/Q)]^{-1}\) is the mechanical susceptibility.
%and \(m_{\mathrm{eff}}=\rho h L^{2}/4\) is the effective mass of a square membrane. 
{The calculated thermomechanical noise agrees well with the measured spectrum in Fig.~\ref{fig:sensitivity}a.}
%The calculated thermomechanical noise spectrum is plotted in Fig.~\ref{fig:sensitivity}a and shows good agreement with the measured noise floor, indicating that the sensor sensitivity is limited by thermomechanical noise. 
%The small discrepancy between theory and experiment is attributed to slow resonance-frequency drift and the finite frequency resolution of the measurement.

Dividing the displacement noise spectrum by the mechanical susceptibility yields the equivalent force noise spectral density, which can be converted into a magnetic field sensitivity via $\sqrt{S_{TT}}={\sqrt{S_{FF}}}/{c_{\mathrm{act}}},$
where \(c_{\mathrm{act}}\) denotes the magnetic actuation coefficient. By comparing the measured sensitivity in Fig.~\ref{fig:step response}d and the thermomechanical force density, we obtain \(c_{\mathrm{act}}\approx \SI{2.86e-4}{N/T}\) and plot the resulting magnetic field noise spectrum in the right $y-$axis of Fig.~\ref{fig:sensitivity}b. As expected, the best force and magnetic field sensitivity is achieved at the mechanical resonance frequency. Under thermomechanical-noise-limited conditions, the minimum detectable magnetic field {amplitude} is therefore given by:
\begin{equation}
  \sqrt{S_{TT}}=\frac{\sqrt{4k_{\mathrm{B}}T m_{\mathrm{eff}}/Q}}{c_{\mathrm{act}}},  
\end{equation}
highlighting the central role of the mechanical quality factor in determining sensitivity.

\section{Discussion and conclusions}

Further reductions in mechanical dissipation could substantially improve the sensitivity of the present platform since it is limited by thermomechanical noise. Quality factors exceeding \(10^{9}\) have been demonstrated in related nanomechanical systems~\cite{cupertino2024centimeter,engelsen2024ultrahigh}. Incorporating resonators with comparable quality factors into our platform is projected to enable sensitivities on the order of \SI{10}{fT/\sqrt{Hz}}, as shown in Fig.~\ref{fig:sensitivity}b. Such sensitivity would be comparable to that of state-of-the-art atomic magnetometers~\cite{kominis2003subfemtotesla,shah2013compact} and superconducting quantum interference devices (SQUIDs)~\cite{weinstock2012squid,drung2007highly} (see also Section SI-X for a comparison with other magnetic field sensing technologies). Unlike atomic magnetometers and SQUIDs, the present levitated nanomechanical magnetometer operates at room temperature and under Earth’s magnetic field. Although the highest sensitivity is achieved near the mechanical resonance, the resonator geometry can be tailored independently to target application-specific operating frequencies.

Although diamagnetic levitation is not a prerequisite for realizing such a modular architecture, it provides a simple, passive, and contact-free implementation and may also open a promising route toward future devices with reduced clamping-induced dissipation and substrate-mediated coupling~\cite{schmid2011damping,de2022mechanical}. More broadly, the hybrid architecture establishes a modular strategy for precision sensing in which magnetic transduction and low-loss mechanical amplification are performed by physically distinct elements. This physical separation avoids direct magnetic functionalization of the ultrahigh-\(Q\) resonator and enables the transducer and resonator to be optimized independently.

In summary, we demonstrate a levitated nanomechanical magnetometer operating at room temperature and without magnetic shielding, achieving a magnetic field sensitivity of \SI{4.5}{pT/\sqrt{Hz}} {at \SI{443}{kHz}}. The centimeter-scale diamagnetically levitated graphite plate provides strong coupling to external magnetic fields, while the high-\(Q\) nanomechanical resonator enables resonant enhancement of the induced motion. The hybrid architecture addresses a central challenge in extending high-$Q$ nanomechanical resonators to broader sensing applications, i.e., the need for efficient coupling to external signals without compromising mechanical dissipation. With further advances in dissipation engineering, this hybrid levitated platform offers a pathway toward femtotesla-level magnetic field sensing under ambient conditions.

\vspace{1cm}
\section*{Acknowledgments}
This research/project is supported by the MTC Individual Research Grants (IRG grant No. M22K2c0090), MTC Young Individual Research Grants (YIRG grant No. M25N8c0141), and the National Research Foundation, Singapore, under its Competitive Research Programme (CRP Award No: NRF-CRP30-2023-0002). Any opinions, findings and conclusions or recommendations expressed in this material are those of the author(s) and do not reflect the views of National Research Foundation, Singapore.

\vspace{1cm}
\section*{Author contributions}
X.C., L.D., and P.K.L. conceived the project. X.C., and N.R. performed the measurements. M.R.C., Y.C. and L.D. fabricated the nanomechanical resonators. X.C., N.R., C.G., Y.C., and S.M.A constructed the measurement setup. X.C. performed the theoretical modeling and numerical simulations. All authors analyzed and interpreted the results. X.C., L.D., and P.K.L secured the funding and supervised the project. X.C. wrote the manuscript with input from all authors. All authors have read and approved the final manuscript.

%TC:ignore
\bibliography{main}% Produces the bibliography via BibTeX.
%TC:endignore

\onecolumngrid
\vspace{1cm}
\begin{center}

\large \textbf{Supplementary Information}
    
\end{center}
\vspace{0.5cm}

\setcounter{section}{0}

\section{Fabrication of samples}

The free-standing high-stress membranes are fabricated from the stoichiometric $\mathrm{Si}_3\mathrm{N}_4$ wafer. The thickness of the $\mathrm{Si}_3\mathrm{N}_4$
layer is \SI{100}{nm} for both sides of the wafer, and the bulk of the wafer is $<100>$ oriented silicon (525 ± \SI{25}{\um} thick). The wafer is diced into $15\times\SI{15}{mm^2}$ coupons for processing. The fabrication procedure is as follows: 
$\mathrm{Si}_3\mathrm{N}_4$/silicon (100) chips are cleaned by a standard procedure to remove organic and ionic contaminants (step 1). After cleaning, one surface of the chip is spin-coated with photoresist (step 2) and membrane patterns are defined on the back surface by electron beam lithography using ZEP (steps 3/4). The patterns are then transferred to the $\mathrm{Si}_3\mathrm{N}_4$ by reactive ion etch (RIE) (step 5). The resist on both surfaces is stripped off, and the exposed silicon is etched in a KOH solution. The KOH etch stops when it reaches the $\mathrm{Si}$ on the front side of the chip, creating membranes. After the KOH etch, the chip with the membranes is rinsed multiple times with deionized water, and followed by an IPA rinse, and then dried (step 6). A microscopic image of the fabricated sample is shown in Fig. \ref{fig:setup-si}b.
\begin{figure}[!h]
\centering
\includegraphics[width=14cm]{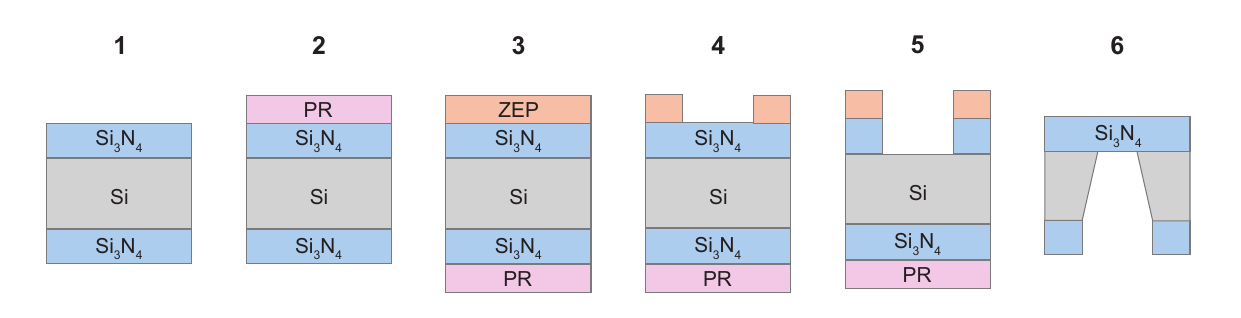}
\caption{\label{fig:concept} $\mathrm{Si}_3\mathrm{N}_4$ membrane fabrication process flow.}
\end{figure}

\section{Derivation of the magnetic force}
In this section, we construct a simple model to calculate the magnetic force applied to the pyrolytic graphite plate. Owing to the strong anisotropy of pyrolytic graphite, the magnetic susceptibility
along the vertical direction is much larger than that in the in-plane directions~\cite{chen2020rigid}. Since only the vertical motion is relevant in our experiments, we restrict the analysis to the vertical direction $y$ and treat all quantities as scalars. Accordingly, the magnetic
field gradient is defined as $\nabla B \equiv \partial B / \partial y$.

The magnetic force per unit volume of the graphite plate along the vertical direction is given by \cite{chen2020rigid}
\begin{equation}
F = \frac{\chi}{\mu_0} B \nabla B ,
\end{equation}
where $B$ denotes the vertical magnetic field component.
We write the magnetic field as the sum of a static magnetic field generated by the permanent
magnets, $B_{\mathrm{m}}$, and a time-dependent signal field, $B_{\mathrm{s}}$,
\begin{equation}
B = B_{\mathrm{m}} + B_{\mathrm{s}} .
\end{equation}
Substituting this expression into the force equation yields
\begin{equation}
F = \frac{\chi}{\mu_0}
\left(
B_{\mathrm{m}} \nabla B_{\mathrm{m}}
+ B_{\mathrm{m}} \nabla B_{\mathrm{s}}
+ B_{\mathrm{s}} \nabla B_{\mathrm{m}}
+ B_{\mathrm{s}} \nabla B_{\mathrm{s}}
\right).
\end{equation}

In our experiments $B_{\mathrm{s}} \ll B_{\mathrm{m}}$, and $B_{\mathrm{m}} \nabla B_{\mathrm{s}}$ is also much smaller than $B_{\mathrm{s}} \nabla B_{\mathrm{m}}$ (see the next section and Fig. \ref{fig:coil-B-F}b). Therefore,  we neglect $B_{\mathrm{m}} \nabla B_{\mathrm{s}}$ and $B_{\mathrm{s}} \nabla B_{\mathrm{s}}$,
obtaining
\begin{equation}
F \approx
\frac{\chi}{\mu_0}
\left(
B_{\mathrm{m}} \nabla B_{\mathrm{m}}
+
B_{\mathrm{s}} \nabla B_{\mathrm{m}}
\right),
\end{equation}
where the first term corresponds to the static levitation force and the second term
represents the signal force responsible for magnetic field sensing. 

Noted that $F_\mathrm{mag}$ is the magnetic force per unit volume of the graphite plate, considering only the $y-$component magnetic field. The total force can be obtained by taking the 3-dimensional magnetic field into account and integrating the unit force over the volume of the graphite plate.

\section{A simple model for the hybrid levitated system}

In this section,  we derive a simple model to describe the dynamics of the hybrid levitated system. The levitated graphite plate is driven by the magnetic force and then its motion is mechanically transferred to the nanomechanical membrane resonator, which acts as a high-\(Q\) mechanical amplifier. The motion of the nanomechanical resonator can be modeled as:
\begin{equation}
    %\left\{
    \begin{aligned}      
    %\Ddot{y}+\frac{\omega_\mathrm{pg}}{Q_\mathrm{pg}}\Dot{y}+\omega_\mathrm{pg}^2 y&=\frac{F_\mathrm{sg}}{m_\mathrm{\mathrm{pg}}}\cos(\omega t), \\
    \Ddot{y}+\frac{\omega_0}{Q}\Dot{y}+\omega_0^2(y-y_\mathrm{g})&=\frac{F_\mathrm{th}}{m_\mathrm{eff}},
    \end{aligned}
    \label{eq:eom-nano}
    %\right.
\end{equation}
%where $m_\mathrm{pg}, Q_\mathrm{pg}$, and $\omega_\mathrm{pg}$ are the effective mass, $Q$-factor, and resonance frequency of the levitated graphite resonator; 
where $y$, $m_\mathrm{eff}$, $Q$, and $\omega_\mathrm{0}$ are the displacement, effective mass, $Q$-factor, and resonance frequency of the nanomechanical resonator; $F_\mathrm{th}$ is the force due to thermomechanical noise.

\textbf{Graphite plate:} The motion of the diamagnetically levitated graphite plate $y_\mathrm{g}$ is determined by the permanent magnets' magnetic field and the total driving force $F_\mathrm{sg}$. The dynamics of the levitated graphite are simplified and modeled as a single-degree-of-freedom oscillator in the z direction. At radio frequencies ($\omega>\SI{100}{kHz}$) that are much higher than the graphite's resonance frequencies, the direct mechanical response of the levitated graphite plate is strongly suppressed, resulting in: 
\begin{equation}
    y_\mathrm{g}=Y_{g}\cos(\omega t)=\frac{F_\mathrm{sg}}{m_\mathrm{pg}\omega^2}\cos(\omega t),    
\end{equation}
where $m_\mathrm{pg}$ is the modal mass of the graphite vibrational mode.

\textbf{Nanomechanical resonator:} The dynamics of a high-stress Si$_3$N$_4$ membrane can be modeled as a driven, damped harmonic oscillator. 
The membrane is square, with side length $L$, thickness $t$, density $\rho$, and intrinsic tensile stress $\sigma_\mathrm{t}$.
In the membrane (tension-dominated) limit, the resonance frequencies of the $(m,n)$ modes are \cite{schmid2016fundamentals}
\begin{equation}
    f_{mn} = \frac{1}{2L}\sqrt{\frac{\sigma_\mathrm{t}}{\rho}}\,\sqrt{m^2 + n^2},  \quad m,n=1, 2, 3, ...
    \label{eq:fmn}
\end{equation}
so that the fundamental $(1,1)$ mode has
\begin{equation}
    f_{11} = \frac{\sqrt{2}}{2L}\sqrt{\frac{\sigma_\mathrm{t}}{\rho}},
    \qquad
    \omega_{11} = 2\pi f_{11}
    = \frac{\pi\sqrt{2}}{L}\sqrt{\frac{\sigma_\mathrm{t}}{\rho}}.
\end{equation}
The frequency depends only on the ratio $\sigma_\mathrm{t}/\rho$ and the lateral size $L$.

The total physical mass of the membrane is
\begin{equation}
    M = \rho t L^2.
\end{equation}
For the fundamental $(1,1)$ mode, the effective modal mass is given by
\begin{equation}
    m_\mathrm{eff} = \frac{M}{4} = \frac{\rho t L^2}{4},
\end{equation}
assuming a readout near the membrane center. 
Therefore, the effective stiffness of the nanomechanical resonator is given by
\begin{equation}
    k_\mathrm{eff} = m_\mathrm{eff}\,\omega_{11}^2 = \frac{\pi^2}{2}\,\sigma_\mathrm{t} t,
\end{equation}
which depends only on the product $\sigma_\mathrm{t} t$, i.e.\ the line tension of the membrane.

In the frequency domain, Eq.~\eqref{eq:eom-nano} yields the mechanical susceptibility
\begin{equation}
    \chi(\omega) = \frac{1}{m_{\mathrm{eff}}(\omega_{0}^{2}-\omega^{2}+i\omega_{0}\omega/Q)}.
\end{equation}
At resonance, $\omega=\omega_0= \omega_{11}$ (for the fundamental mode), the displacement amplitude is maximized,
\begin{equation}
    X_\mathrm{0} 
    = \left|\chi(\omega_{0})\right| F_0
    = \frac{Q}{m_\mathrm{eff}\omega_{0}^2}\,F_0,
\end{equation}
where $F_0$ is the applied force amplitude.
Substituting Eq. (10) and Eq. (8) into Eq. (13) gives a simple expression for the fundamental mode:
\begin{equation}
    X_\mathrm{0} 
    = \frac{2}{\pi^2}\,\frac{Q}{\sigma_\mathrm{t} t}\,F_0.
\end{equation}
When driven by the magnetic signal, $F_0=k_\mathrm{eff}Y_\mathrm{g}=k_\mathrm{eff}\frac{F_\mathrm{sg}}{m_\mathrm{pg}\omega^2}$ from Eq. (6). Together with Eq. (11), we have
\begin{equation}
    X_\mathrm{0} 
    = \frac{2}{\pi^2}\,\frac{Q}{\sigma_\mathrm{t} t}\,\frac{m_\mathrm{eff}}{m_\mathrm{pg}}\,F_\mathrm{sg}= \frac{1}{2\pi^2}\,\frac{Q \rho L^2}{\sigma_\mathrm{t}}\,\frac{F_\mathrm{sg}}{m_\mathrm{pg}}.
\end{equation}
Therefore, a higher $Q$ and larger $L$ result in a more significant response in the nanomechanical resonator at the resonance frequency. In other words, from Eq. (8), a higher $Q$ and smaller $f_\mathrm{11}$ lead to a more significant response.  

\section{Interferometric setup and calibration}

\begin{figure}[!h]
\centering
\includegraphics[width=12cm]{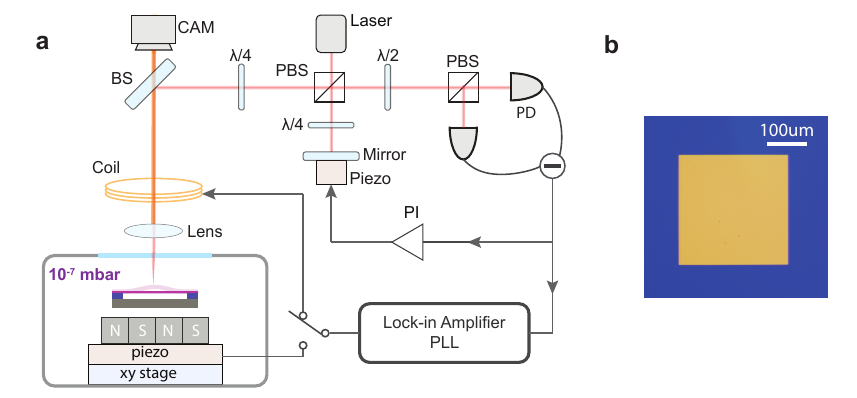}
\caption{\label{fig:setup-si} (a) Optical interferometric readout of the nanomechanical resonator motion. PBS: polarized beam splitter; BS: beam splitter; PD: balanced photo detector; $\lambda/2$: half-wave plate; $\lambda/4$: quarter-wave plate;  CAM: camera; PI: proportional-integral controller; PLL: phase-locked loop control. (b) A microscopic image of a SiN nanomechanical membrane sample.}
\end{figure}

The motion of the nanomechanical resonators is measured using a Michelson
interferometer, as shown in Fig.~\ref{fig:setup-si}a. A coherent laser with wavelength $\lambda=\SI{1550}{nm}$ is used as the light source to minimize
optical absorption in the silicon nitride membranes. The laser beam is split
into a probe arm and a reference arm, recombined after reflection from the
nanomechanical membrane and a reference mirror, respectively, and detected
using a balanced photodetector. The differential signal is then sent to a
Zurich lock-in amplifier for data acquisition and processing.

To suppress slow phase drift of the interferometer, a proportional--integral
(PI) feedback controller implemented on a National Instruments FPGA is used
to stabilize the interferometer at the mid-fringe operating point. The
feedback signal is applied to a piezoelectric actuator mounted on the
reference-arm mirror, ensuring linear transduction of membrane displacement.

\begin{figure}[!h]
\centering
\includegraphics[width=8cm]{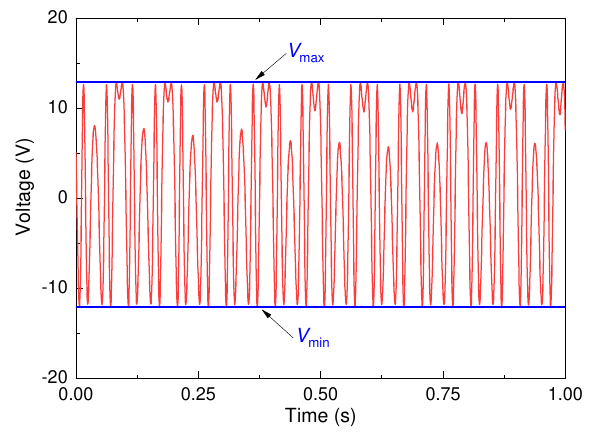}
\caption{\label{fig:inter-cali} Voltage signal as a function of time measured from the balanced detector for displacement calibration.}
\end{figure}

The interferometer is calibrated by applying a known sinusoidal displacement
to the reference-arm piezoelectric actuator. A representative calibration
result is shown in Fig.~\ref{fig:inter-cali}. Since the displacement amplitude
of the nanomechanical resonators is much smaller than the laser wavelength
$\lambda$, the interferometer operates in the linear regime, and the measured
voltage signal can be converted into displacement according to~\cite{barg2017measuring}
\begin{equation}
\delta x(t) \approx \frac{\lambda}{4\pi V_{\mathrm{ff}}}\,\delta V(t),
\end{equation}
where $V_{\mathrm{ff}} = (V_{\mathrm{max}} - V_{\mathrm{min}})/2$ denotes the
full-fringe voltage of the interferometer.

\section{Phase-locked loop measurement}
\begin{figure}[!h]
\centering
\includegraphics[width=14cm]{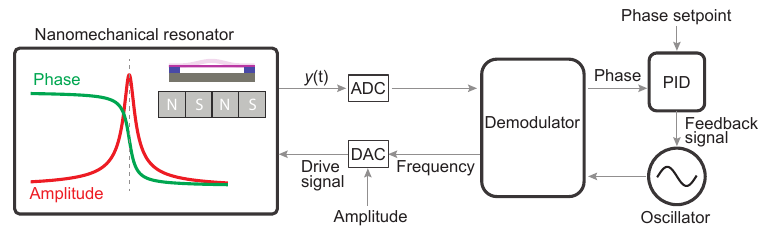}
\caption{\label{fig:PLL} Diagram of the phase-locked loop measurement.}
\end{figure}

A phase-locked loop (PLL) is employed to track the real-time {amplitude} response of the nanomechanical resonator while continuously driving it at its resonance frequency, as shown in Fig.~\ref{fig:PLL}. The PLL is implemented in a closed-loop configuration using a Zurich Instruments UHFLI lock-in amplifier.

Prior to activating the PLL, the resonance frequency, phase response, and quality factor of the resonator are determined by measuring its frequency response function curve. Based on this characterization, the PLL is configured to lock the resonator phase to $\pi/2$ after correcting for the shift introduced by the equipment using a proportional–integral–derivative (PID) controller. As a result, the driving frequency is continuously adjusted to follow the instantaneous resonance frequency of the resonator, ensuring that the system remains driven at resonance throughout the measurement.

\section{Q-factors extracted from ringdown measurements}
The Q-factors of the six devices studied in this work are determined using a ring-down measurement technique. Each resonator is first driven at its resonance frequency and then allowed to freely decay after the excitation is switched off. The subsequent displacement amplitude is recorded using a Zurich lock-in amplifier.

\begin{figure}[h!]
\centering
\includegraphics[width=14cm]{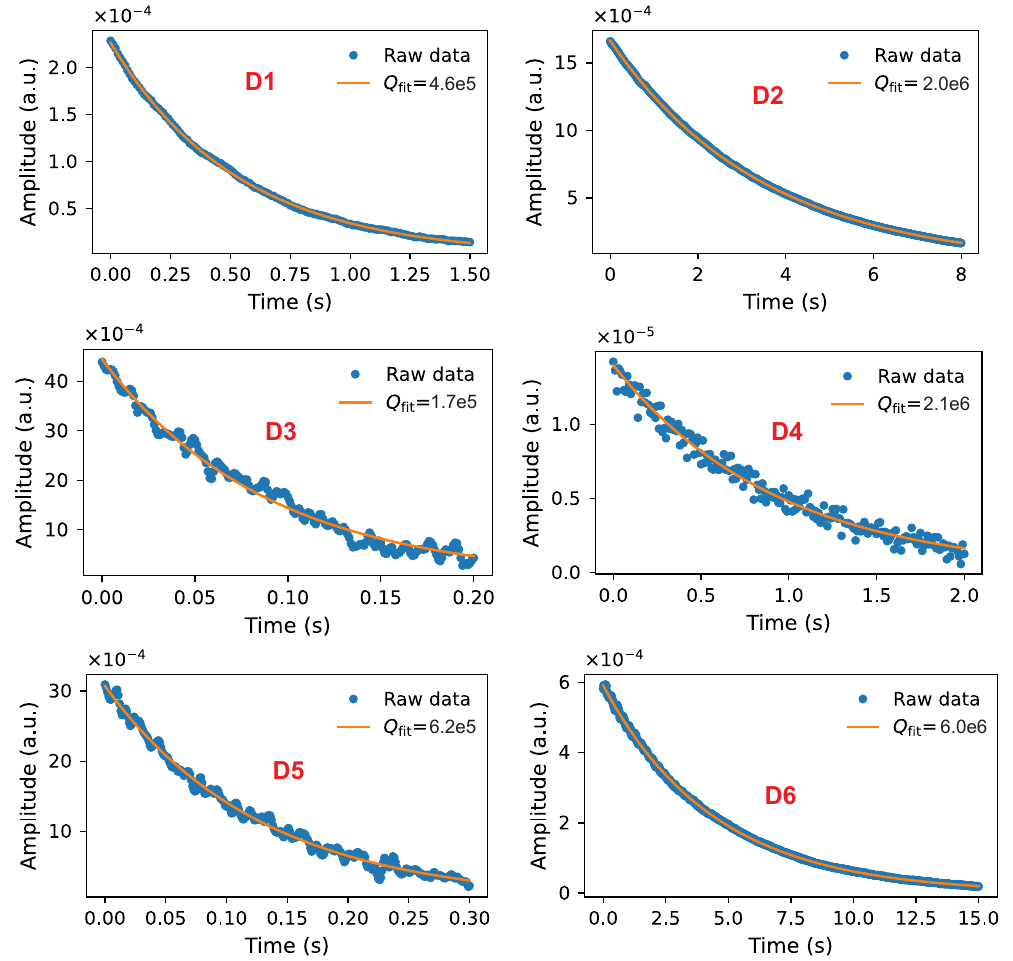}
\caption{\label{fig:ringdown-Q} Ring-down measurement to extract the Q-factors for the 6 devices.}
\end{figure}

For each device, the ring-down measurement is repeated multiple times, and the reported Q-factors are the averaged values extracted from these measurements, as shown in Fig.~\ref{fig:ringdown-Q}. The measurements are performed at different excitation amplitudes, which leads to varying levels of fluctuation in the recorded decay traces but does not affect the extracted quality factors within the experimental uncertainty.

\section{Coil calibration}
\begin{figure}[!h]
\centering
\includegraphics[width=12cm]{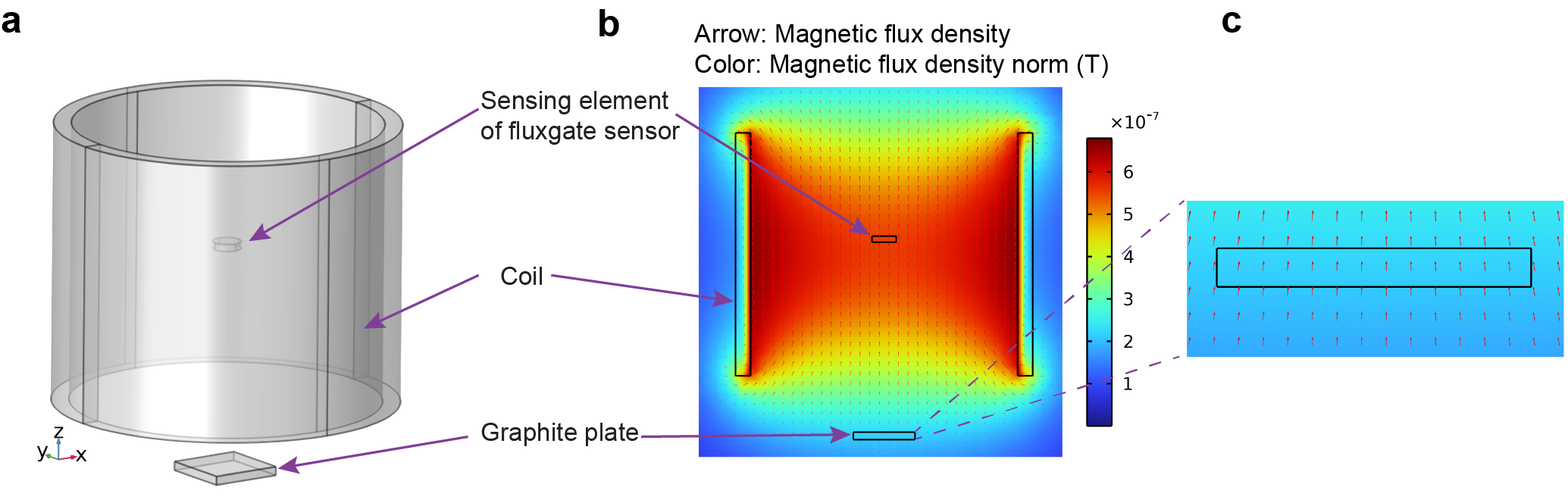}
\caption{\label{fig:coil-cali-geo} (a) Diagram of the coil and the pyrolytic graphite plate. (b) Generated magnetic field from the coil when applied with \SI{0.6}{V}. The color represents the magnetic flux density norm; the arrow represents the magnetic flux density in a 2D surface. (c) Zoom-in to the graphite plate.}
\end{figure}
To calibrate the levitated magnetometer, an electromagnetic coil is used to generate a controllable magnetic field. The coil is fabricated by winding a copper wire (24~AWG) around a cylindrical former with a diameter of \SI{44}{mm} and a height of \SI{40}{mm}, resulting in a nominal total of 310 turns. To limit the current and prevent overheating or short-circuiting under DC operation, a series resistor of \SI{47}{\ohm} is included in the circuit.

\begin{figure}[!h]
\centering
\includegraphics[width=15cm]{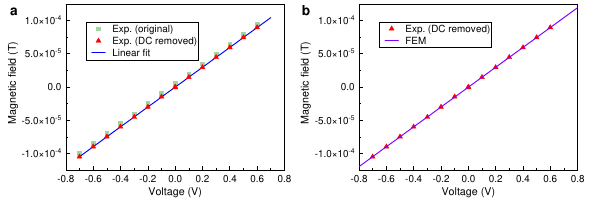}
\caption{\label{fig:coil-cali} Coil calibration with DC voltage. (a) Measured DC magnetic field at the center of the coil when applied with different DC voltages. (b) Comparison between the measured magnetic field with DC offset removed and the one obtained from FEM.}
\end{figure}

The coil is first calibrated in the DC regime using a commercial fluxgate magnetometer. The sensing element of the fluxgate sensor is placed at the center of the coil and aligned with the vertical (\(z\)) axis of the coil, as shown in Fig.~\ref{fig:coil-cali-geo}a. The measured magnetic field along the \(z\) direction for applied voltages ranging from \SI{-0.8}{V} to \SI{0.8}{V} is shown in Fig.~\ref{fig:coil-cali}a. An offset magnetic field is observed at zero applied voltage, which is attributed to ambient magnetic fields. After subtracting this DC offset, the data are well described by a linear fit, confirming a linear relationship between the generated magnetic field and the applied voltage.

To benchmark the experimental calibration, finite-element simulations are
performed using COMSOL Multiphysics. The coil is modeled using its physical dimensions and electrical parameters, including the series resistor. Figure~\ref{fig:coil-cali-geo}b shows the simulated magnetic field distribution on a cross section of the coil, where the color scale represents the magnetic flux density magnitude and the arrows indicate the magnetic flux density direction. As expected, the magnetic field reaches its maximum at the center of the coil. A zoomed-in view of the magnetic field in the vicinity of the levitated graphite plate further confirms that the magnetic field amplitude varies only weakly across the plate area.

To enable a direct comparison with the experimental measurements, the
simulated \(z\)-component of the magnetic field is averaged over the sensing volume of the fluxgate sensor (indicated in Fig.~\ref{fig:coil-cali-geo}a,b). The resulting simulated values are shown in Fig.~\ref{fig:coil-cali}b and exhibit good agreement with the measured data.
In the simulations, a coil filling factor of 0.84 is used to account for the finite packing density of the windings, which is obtained by fitting the experimental data.

\begin{figure}[!h]
\centering
\includegraphics[width=15cm]{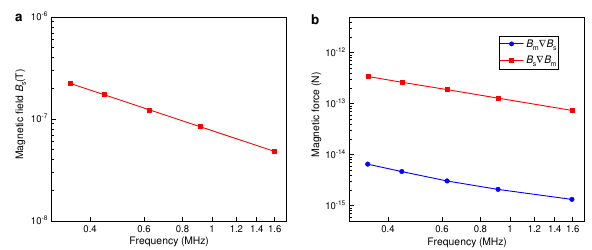}
\caption{\label{fig:coil-B-F} (a) Simulated magnetic field {amplitude} at the center of the coil when applied with different frequencies. (b) Calculated magnetic force {amplitudes} arising from \(B_{\mathrm{m}} \nabla
B_{\mathrm{s}}\) and \(B_{\mathrm{s}} \nabla B_{\mathrm{m}}\), respectively.}
\end{figure}

After calibrating the coil in the DC regime, the same model is used to
calculate the magnetic field response at higher frequencies.
Figure~\ref{fig:coil-B-F}a shows the calculated magnetic field amplitude as a function of frequency for an applied voltage {amplitude} of \SI{0.6}{V}. Since the magnetic field varies only weakly over the spatial extent of the levitated graphite plate (see Fig.~\ref{fig:coil-cali-geo}c), the \(z\)-component of the magnetic field {amplitude} is spatially averaged over the plate volume. This allows us to establish the relationship between the applied voltage {amplitude} and the generated magnetic field {amplitude} over a wide frequency range, and to benchmark the performance of our levitated magnetometer.

Using the model, we further evaluate the magnetic
forces arising from the two cross terms \(B_{\mathrm{m}} \nabla
B_{\mathrm{s}}\) and \(B_{\mathrm{s}} \nabla B_{\mathrm{m}}\). In COMSOL, the magnetic field {amplitude} generated by the coil is simulated using the Magnetic Fields and Electrical Circuit interfaces (Study~1), while the magnetic field generated by the permanent magnets is simulated using the Magnetic Fields, No Currents interface (Study~2). In the post-processing stage, the force contributions \(\chi/\mu_0\, B_{\mathrm{m}} \nabla B_{\mathrm{s}}\) and \(\chi/\mu_0\, B_{\mathrm{s}} \nabla B_{\mathrm{m}}\) are integrated over the volume of the pyrolytic graphite plate.

The results, shown in Fig.~\ref{fig:coil-B-F}b, demonstrate that the force term \(B_{\mathrm{s}} \nabla B_{\mathrm{m}}\) is approximately two orders of magnitude larger than \(B_{\mathrm{m}} \nabla B_{\mathrm{s}}\) for the experimental parameters used in this work. This confirms that the dominant signal force originates from the interaction between the applied magnetic field amplitude and the static magnetic field gradient provided by the permanent magnets. Consequently, the contribution from the magnetic field gradient generated by the coil is neglected in the present analysis.

\section{Allan deviation analysis}

To quantify the stability of the levitated magnetometer, we compute the Allan deviation of the displacement time series $x_i$ of the nanomechanical resonator while it is driven at its resonance frequency continuously using the phase-locked loop with different excitation voltage {amplitudes}.

The time-series data are divided into blocks characterized by an averaging (or gate) time $\tau = N\,\Delta t$, where $\Delta t$ is the sampling period and $N$ is an integer number of samples. Specifically, the data are partitioned into successive, non-overlapping blocks of length $N$.

The mean displacement within the $k$-th block is given by
\begin{equation}
\bar{x}_k(\tau)
= \frac{1}{N} \sum_{i = k N}^{(k+1)N - 1} x_i ,
\qquad
k = 0, \dots, K-1 ,
\end{equation}
where $N_\mathbf{tot}$ is the total number of samples and
$K = \lfloor N_\mathbf{tot}/N \rfloor$ is the total number of complete blocks.

The Allan variance is then defined as
\begin{equation}
\sigma_{\mathrm{a}}^2(\tau)
= \frac{1}{2(K-1)}
\sum_{k=0}^{K-2}
\left[
\bar{x}_{k+1}(\tau) - \bar{x}_k(\tau)
\right]^2 ,
\end{equation}
and the Allan deviation is given by
$\sigma_{\mathrm{a}}(\tau) = \sqrt{\sigma_{\mathrm{a}}^2(\tau)}$.

In our analysis, the averaging time $\tau$ is swept logarithmically from $\tau = \SI{0.01}{s}$ up to
\SI{20}{s}. The lower limit is set by the low-pass filter of our lockin amplifier setting and the upper limit ensures a sufficient number of independent samples for reliable statistical estimation at long averaging times.

%{\color{red}
\section{Influence from the magnets and supporting structures}
\begin{figure}[]
\centering
\includegraphics[width=6cm]{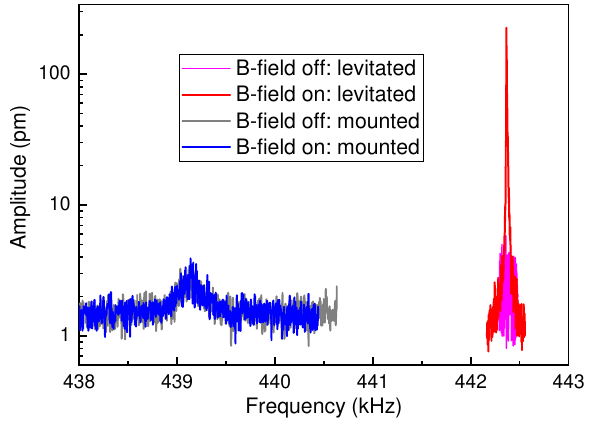}
\caption{\label{fig:control} Frequency response curves of a membrane under different mounting conditions when the magnetic field from a coil is on and off, respectively.}
\end{figure}

To support that the measured membrane response originates from the levitated graphite rather than from the magnets or the supporting structure, we performed a control experiment by varying a membrane's mounting conditions. Specifically, the membrane chip was directly mounted onto the permanent magnets. Figure~\ref{fig:control} compares the measured nanomechanical resonator response around its first resonance frequency under levitated and directly mounted conditions.

A pronounced resonant response is clearly observed in the levitated configuration, whereas no comparable driven response is observed when the membrane chip is directly mounted onto the permanent magnets. The small remaining peak in the mounted configuration is consistent with the thermomechanical noise of the nanomechanical resonator. The resonance frequency of the mounted membrane shifts slightly, likely due to the change in mounting conditions. In addition, the frequency ranges and measurement points/durations are different in the two measurements, causing the spectrum to be visually different. These results confirm that the measured driven response predominantly originates from the levitated graphite transducer, while contributions from the magnets and supporting structure are negligible under properly fixed conditions.

In addition, even if a small residual motion of the magnets or support structure were present, its transfer to the levitated graphite would be strongly suppressed at the frequencies relevant to the nanomechanical resonator ($>\SI{100}{kHz}$). This is because the levitated graphite behaves as a base-excited single-degree-of-freedom mechanical resonator with a much lower resonance frequency if the magnets are driving the levitated graphite. The vertical resonance frequency of the levitated graphite is $f_\mathrm{gra}=\SI{25.4}{Hz}$, while the measured magnet motion above \SI{100}{kHz} is below $y_\mathrm{mag}=\SI{0.1}{pm}$ when the magnetic field is switched on. For a base-excited single-degree-of-freedom resonator, the absolute motion of the resonator is suppressed relative to the base motion by approximately $(f_\mathrm{gra}/f)^2$ in the high-frequency limit ($f\gg f_\mathrm{gra}$). Therefore,
\[
x_\mathrm{gra}\simeq y_\mathrm{mag}\left(\frac{f_\mathrm{gra}}{f}\right)^2.
\]
At $f=\SI{100}{kHz}$, the induced absolute motion of the graphite is estimated to be
\[
x_\mathrm{gra}\lesssim \SI{6.5e-9}{pm}.
\]
This value is negligibly small, indicating that residual motion of the magnets or support structure cannot account for the observed driven membrane response.

\section{Comparison with other magnetic field sensing technologies}

\vspace*{2.5mm}
\noindent
In this section, we add a comparison table (Table~\ref{tab:magnetometer_comparison}) to benchmark our work against representative state-of-the-art magnetic field sensing technologies, with a particular focus on optomechanical and nanomechanical magnetometers.

From our literature survey, we observe that although substantial progress has been made in developing ultrahigh-$Q$ nanomechanical resonators~\cite{engelsen2024ultrahigh}, there are still relatively few demonstrations of magnetic field sensors based on such resonators. One important challenge is the integration of magnetic functionality while preserving the ultrahigh mechanical quality factor. As summarized in Table~\ref{tab:magnetometer_comparison}, only one previous work has demonstrated a magnetometer based on a high-$Q$ SiN membrane resonator by integrating Terfenol-D plates into the device structure, achieving a sensitivity approximately one order of magnitude worse than our work.

Over the past decade, a major research direction has focused on optomechanical magnetometers based on magnetostrictive materials integrated with optical cavities, with recent demonstrations achieving sub-picotesla sensitivity. These magnetometers primarily rely on ultrahigh-sensitivity optical cavity readout and strong optomechanical coupling to enhance the magnetic response. In contrast, our work proposes a different strategy by physically separating the magnetic transduction component from the ultrahigh-$Q$ nanomechanical resonator. This architecture avoids introducing magnetic materials directly onto the resonator and therefore preserves the exceptionally high mechanical quality factor of the SiN membrane.

In addition, compared with technologies such as SERF atomic magnetometers and SQUIDs, our approach offers several potential advantages, including room-temperature operation, compact solid-state implementation, and operation at much higher frequencies. While the current sensitivity of our device does not yet reach the level of SERF or SQUID magnetometers, our results demonstrate a promising route toward highly sensitive solid-state magnetometry based on ultrahigh-$Q$ nanomechanical resonators. Considering that state-of-the-art nanomechanical resonators can now achieve quality factors exceeding $10^9$ \cite{engelsen2024ultrahigh}, our approach has strong potential to further improve toward the femto-Tesla regime.

\begin{table*}[t]
\caption{Comparison of magnetic field sensing technologies.}
\label{tab:magnetometer_comparison}
\centering
\scriptsize
\renewcommand{\arraystretch}{1.55}
\setlength{\tabcolsep}{3pt}

\begin{tabular}{lllllll}
\hline\hline
\textbf{Technology} &
\textbf{Type} &
\textbf{Sensitivity} &
\textbf{Frequency} &
\textbf{Q} &
\textbf{Note} &
\textbf{Ref.} \\
& & \textbf{(T/$\sqrt{\mathrm{Hz}}$)} & \textbf{(Hz)} & & & \\
\hline

\raisebox{-0.4cm}[0pt][0pt]{\parbox{1.8cm}{\centering High-Q SiN\\nanomechanical\\resonator}}
& \parbox{5.2cm}{\textbf{Levitated high-Q SiN membrane + interferometer}}
& \textbf{4.50E-12}
& \textbf{3.40E+05}
& \textbf{6.00E+06}
& \textbf{RT}
& \textbf{This work} \\[1.0ex]

&
\parbox{5.2cm}{High-Q SiN membrane + Terfenol-D plates}
& 1.79E-11
& 3.80E+05
& 9.50E+04
& RT
& \cite{zhang2025membrane} \\[1.0ex]

\hline

\raisebox{-2.0cm}[0pt][0pt]{\parbox{1.8cm}{\centering Optomechanical}}
& \parbox{5.2cm}{Magnetostrictive gap-swing Fabry--Perot cavity}
& 6.20E-13
& 5.00E+03
& $\sim$60
& RT
& \cite{xu2024subpicotesla} \\[1.0ex]

&
\parbox{5.2cm}{FeGaB-coated WGM microdisk}
& 1.68E-12
& 9.52E+06
& $\sim$
& RT
& \cite{hu2024picotesla} \\[1.0ex]

&
\parbox{5.2cm}{Terfenol-D microtoroid cavity}
& 4.00E-07
& $\sim$1E+07
& $\sim$1000
& RT
& \cite{forstner2012cavity} \\[1.0ex]

&
\parbox{5.2cm}{Macroscopic CaF$_2$ WGM resonator}
& 1.31E-10
& 1.27E+05
& $\sim$
& RT
& \cite{yu2016optomechanical} \\[1.0ex]

&
\parbox{5.2cm}{Squeezed-light microtoroid magnetometer}
& $\sim$5E-9
& 8.54E+06
& $\sim$
& RT
& \cite{li2018quantum} \\[1.0ex]

&
\parbox{5.2cm}{Terfenol-D microtoroid microcavity}
& 2.60E-11
& 1.05E+07
& $\sim$
& RT
& \cite{li2020ultrabroadband} \\[1.0ex]

&
\parbox{5.2cm}{YIG FMR + glass microsphere}
& 8.50E-10
& 2.06E+08
& $\sim$
& RT
& \cite{arregui2020ferromagnetic} \\[1.0ex]

\hline

SERF
& \parbox{5.2cm}{Spin-Exchange Relaxation-Free magnetometer}
& 5.00E-16
& 10--150
& --
& HT
& \cite{kominis2003subfemtotesla} \\[1.0ex]

SQUID
& \parbox{5.2cm}{Superconducting Quantum Interference Device}
& 3.60E-15
& DC--MHz
& --
& CT
& \cite{drung2007highly} \\[1.0ex]

MEMS
& \parbox{5.2cm}{Microelectromechanical systems}
& $\sim$1E-9
& 1k--100k
& --
& RT
& \cite{herrera2009resonant} \\[1.0ex]

\hline\hline
\end{tabular}

\vspace{1mm}
{\scriptsize RT: room temperature; HT: high temperature; CT: cryogenic temperature.}

\end{table*}

\end{document}